\documentclass[namedreferences,hyperref,optionalrh,solaromanenum]{spr-sola}

\usepackage{graphicx}                    % For eps figures, newer & more powerfull
\usepackage{color}                       % For color text: \color command
\usepackage{booktabs,caption}
\usepackage[flushleft]{threeparttable}
\usepackage{bm}
\usepackage{adjustbox}
\usepackage{longtable}
\usepackage[labelsep=space]{caption}
\newcommand{\etal}{et al.}

\chardef\us=`\_

\begin{document}

\begin{frontmatter}

\title{Doppler Velocity Variation near the Solar Limb in the Solar Photosphere Observed with Hinode}

%%%%%%%%%%%%%%%%%%%%%%%%%%%%%%%%%%%%%%%%%%%%%%%%%%%
%% Authors Names
%
% \author[addressref={},corref,email={}]{\inits{}\fnm{}\snm{}\orcid{}}
\author[addressref={aff1,aff2},corref,email={akie.moritsuka@grad.nao.ac.jp}]{\inits{A.}\fnm{Akie}~\lnm{Moritsuka}}
\author[addressref={aff2,aff1},corref,email={yukio.katsukawa@nao.ac.jp}]{\inits{Y.}\fnm{Yukio}~\lnm{Katsukawa}\orcid{0000-0002-5054-8782}}
\author[addressref={aff3,aff4},corref,email={ishikawa.ryohtaro@nifs.ac.jp}]{\inits{R.T.}\fnm{Ryohtaroh T.}~\lnm{Ishikawa}\orcid{0000-0002-4669-5376}}

%%%%%%%%%%%%%%%%%%%%%%%%%%%%%%%%%%%%%%%%%%%%%%%%%%%
%% Runningheads
%
%\runningauthor{}
%\runningtitle{}
\runningauthor{A. Moritsuka \etal}
\runningtitle{Doppler Velocity Variation near the Solar Limb}

%%%%%%%%%%%%%%%%%%%%%%%%%%%%%%%%%%%%%%%%%%%%%%%%%%%
%% Affilations 
%% id shold be the same with \author addressref value.
%\address[id={}]{}
\address[id=aff1]{Department of Astronomy, Graduate School of Science, The University of Tokyo, 7-3-1 Hongo, Bunkyo-ku, Tokyo 113-0033, Japan}
\address[id=aff2]{National Astronomical Observatory of Japan, 2-21-1 Osawa, Mitaka, Tokyo 181-8588, Japan}
\address[id=aff3]{National Institute for Fusion Science, 322-6 Oroshi-cho, Toki, Gihu 509-5292, Japan}
\address[id=aff4]{Fusion Science Program, Graduate Institute for Advanced Studies, The Graduate University for Advanced Studies, SOKENDAI, 322-6 Oroshi-cho, Toki, Gifu 509-5292, Japan}

%%%%%%%%%%%%%%%%%%%%%%%%%%%%%%%%%%%%%%%%%%%%%%%%%%%
%%% Abstract 
\begin{abstract}
The Doppler velocity and bisector of photospheric lines show significant center-to-limb variation. This variation has been considered to be caused by the projection angle of the granular convection which has a corrugated surface. This study aimed to show the spectral line shifts up to limb and investigate how these shifts are caused by granular convection. We analyzed the Fe~{\sc i}~$630.15$~{nm} absorption line by spatially resolving the granulation very close to the limb using datasets of polar observations obtained with the Spectro-polarimeter (SP) of the Solar Optical Telescope (SOT) onboard Hinode. We also analyzed the Fe~{\sc i}~$630.15$~{nm} emission line, which was observed within $1^{\prime\prime}$ above the limb. This study revealed the following: 1) the spatially averaged spectral lines showed an increase in redshift toward the limb, then became constant, and finally became nearly zero, 2) the Doppler velocities of granules, which were classified as bright and non-magnetic regions, showed as increase in redshifts toward the limb, 3) the bisector of granules changed more than that of the intergranules, and 4) the emission lines showed small blueshifts. These results indicate that the increase in redshift toward the limb is caused by the increase in the redshift of granules. Because the bisector reflects the atmospheric velocity along the line-of-sight, our observations can be interpreted as suggesting that the line-forming region in granules is geometrically thicker than that in intergranules.
\end{abstract}

%%%%%%%%%%%%%%%%%%%%%%%%%%%%%%%%%%%%%%%%%%%%%%%%%%%
%% Keywords
%

\end{frontmatter}
%-------------------------------------------------

%%%%%%%%%%%%%%%%%%%%%%%%%%%%%%%%%%%%%%%%%%%%%%%%%%%
%% Sections
%
\section{Introduction}\label{section:introduction} 
The solar photosphere is covered with a convective pattern known as granulation; the bright and hot upward flow is called the granules, and the dark and cold downward flow surrounding the granules is called the intergranules \citep{dravins1975physical, dravins1981solar}. Because of the correlation between brightness and velocity, the spatially averaged spectral line of the photosphere shows a blueshift at the disk center. This shift due to convection is called the convective blueshift. The magnitude of the convective blueshift is different for each spectral line \citep{appenzeller1967center, dravins1981solar}. The Doppler velocity of the linecore depends on the depth of the spectral lines: the cores of shallow spectral lines are blueshifted more than those of deep lines \citep{ellwarth2023convective}. The shifts of photospheric lines show significant center-to-limb variation \citep{balthasar1988the}. The blueshift observed at the disk center decreases toward the limb, and a slight redshift is observed near the limb. Some spectral lines show an initial increase in blueshift away from the disk center, around $\cos\theta = 0.8$ where $\theta$ is the heliocentric angle, and this increase in blueshift is caused by the horizontal flow of granulation  \citep{lohner2018convective, lohner2019convective, Stief2019convective}. 

The photospheric lines are known to exhibit asymmetry in their profiles. This asymmetry is detected using the bisector method, where the bisector is defined as the midpoint of a line segment at a certain depth of a single absorption profile. Because the line core is formed in the upper layers of the photosphere and the wings are formed in the lower layers, the bisector indicates the physical conditions at different formation heights. Thus, the bisector method has been used to study the line-of-sight variation of the Doppler velocity. Through a three-dimensional (3D) radiative magnetohydrodynamics (MHD) simulation, \citet{manrique2020Capabilities} demonstrated that the bisector velocities are correlated with the line-of-sight velocities. The bisector of spatially averaged spectral lines exhibits a C-shape around the disk center and changes to a \textbackslash-shape toward the limb  \citep{adam1976the, balthasar1984asymmetries, ellwarth2023IAG}. In spatially resolved observations, locally enhanced line asymmetries caused by granular dynamics have been reported (e.g. \citealp{ishikawa2020study}).

These center-to-limb variations of the Doppler velocities and the bisector are known as the limb effect. The limb effect is known to originate from convection because it is not observed in sunspots, where convection is suppressed \citep{beckers1977material}. The redshift near the limb has been understood to be caused by the corrugated surface of the photosphere and an increase in formation height. The redshifted horizontal flow of the front granule appears brighter than the blueshifted horizontal flow because the sides of the back granule provide the background light \citep{beckers1978some, cegla2018stellar}. When the front granule conceals the back granule, the blueshifted horizontal flow of the back granule is hidden, and the redshifted flow is more observed \citep{balthasar1985on}. Additionally, the bright wall above the intergranules hides the blueshifted flow of the back granule \citep{delacruzrodriguez2011solar}.

The Doppler shifts and line asymmetries up to the solar limb are not well revealed. Performing spectroscopic observations very close to the solar limb, along with a high spatial resolution, is challenging. Moreover, a long ray path is required for numerical synthesis of the spectral lines very close to the solar limb. In particular, quantitative explanations for the effects of the corrugated 3D convection in the photosphere are still subject to debate. The aim of this study was to quantify the effect of granular convection on spectral lines very close to the limb with high resolution by the Solar Optical Telescope (SOT; \citealp{tsuneta2008solar}) onboard Hinode \citep{kosugi2007hinode}. We investigated the Doppler velocity and bisector very close to the limb, which has not been previously clarified. We classified the observed spectra into those from granules, intergranules, and magnetic regions. \citet{lites2010scattering} revealed the existence of a thin layer in which the Fe~{\sc i}~$630.15$~nm line can be observed as an emission line. We use both absorption and emission lines observed with the Spectro-Polarimeter (SP) of the Hinode SOT. \cite{shelyag2015spectro} reproduced the emission profiles using the numerical simulation code MURaM, suggesting that the transition of absorption-emission profiles at the limb is caused by localized temperature increases and torsional flows within magnetic flux tubes.

In Sect. 2, we describe observations used in our analysis. In Sect. 3, we show the analysis method and the classification of the solar atmosphere by continuum intensities and total polarization degrees. In Sect. 4, we show the center-to-limb variation of the Doppler velocity and bisector velocities. In Sect. 5, we summarize the observation results and discuss the three-dimensional structure of the solar surface inferred from the Doppler velocities and bisectors near the limb.

\section{Observations}\label{section:observations} 
We used the spectropolarimetric data taken with SOT onboard the Hinode satellite. The SOT was a $50$~{cm} aperture telescope with a diffraction limit resolution of $0^{\prime\prime}.31$ at $630$~{nm}. The SOT/SP obtained Stokes profiles (I, Q, U, V) which covered Fe~{\sc i}~$630.15$~{nm} and Fe~{\sc i}~$630.25$~{nm} lines with a spectral sampling of 21.5~{m\AA}. The Fe~{\sc i}~lines were observed as the absorption line on the disk, and as the emission line within $1^{\prime\prime}$ above the limb. A two-dimensional spectral map was obtained by slit-scanning in the east-west direction. The spatial resolution in the slit direction (north-south) was $0^{\prime\prime}.16$ and in the scan direction (east-west) was $0^{\prime\prime}.15$. We used the Level-1 calibrated data processed with the standard calibration routine SP\_PREP \citep{lites2013spprep}. We primarily used the Fe~{\sc i}~$630.15$~{nm} line because the line was less sensitive to the Zeeman effect than the Fe~{\sc i }~$630.25$~{nm} line. 

In this study, we used $49$ datasets observed with the Hinode SOT/SP to investigate the Doppler velocities of the photosphere very close to the limb. These datasets were selected to satisfy the following two requirements. The first requirement was that either the north or south limb was included in the field-of-view. The second requirement was that the datasets were taken with the best spatial resolution without binning. Table \ref{table:datasets} lists the $49$ datasets which we use in this study. $N_x$ is the number of slit positions. $N_y$ is the number of pixels in the north-south direction and is $1024$ for all datasets. $n_x$ is the number of slit positions used in the analysis. The tip-tilt mirror \citep{shimizu2008image} was sometimes reset during scanning, and some data had shifts in the north-south direction of the field-of-view. We visually checked the continuum map of all datasets and cut out the areas without shifts for our analysis. 

Because we required an accurate limb position to obtain the heliocentric angle dependence of the Doppler velocities very close to the limb, the pixel with the largest gradient of the continuum intensity was detected as the limb position at each slit position. After detecting the limb position at all slit positions, we performed the circle fitting on the limb to obtain the coordinates of the disk center. Finally, we determined the heliocentric angle from the disk center for each spatial pixel within the disk.

\begin{figure}[htbp]    %%%%%%%%%%%%%%%%%% MAP
   \centerline{\includegraphics[width=1.0\textwidth,clip]{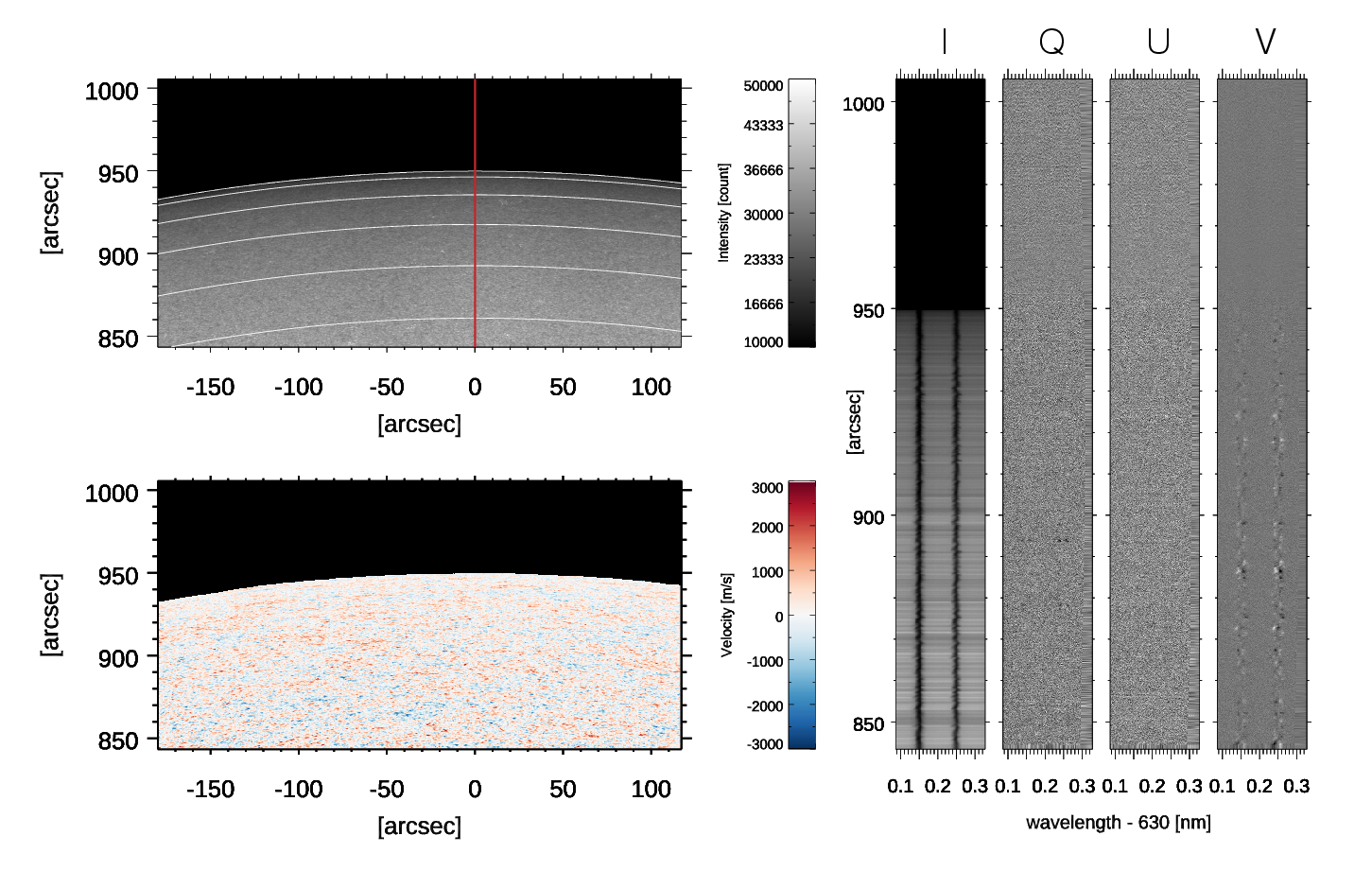}}
 \caption{Left: Maps of continuum intensity (top) and Doppler velocities (bottom)  in the north polar region taken with the Hinode SOT/SP on 2007 September 1 (dataset No. 13 in Table \ref{table:datasets}). The horizontal and vertical axes show the distance from the disk center in arcsec. The contour lines show heliocentric angles $\theta = 65, 70, 75, 80, 85,$ and $89^{\circ}$. The red line indicates the $x=0$ position. The Doppler velocity in the bottom panel is the average of the Doppler velocities of the line core and the lower 5\% of the bisector. Right: Stokes I, Q, U, V spectrum at $x=0$, which is shown by the red line in the upper left panel.  See Figure \ref{figure:map2} for another dataset taken in the south polar region.
}
    \label{figure:map}
 \end{figure}

The left panels of Figure \ref{figure:map} show the continuum and Doppler images taken on 2007 September 6 from 00:28 to 05:57 UT with the Hinode SOT in the north polar region (dataset No. 13 in Table \ref{table:datasets}). The contour lines in the upper panel show the heliocentric angles at $\theta = 65, 70, 75, 80, 85,$ and $90^{\circ}$. The right panels of Figure \ref{figure:map} show the Stokes IQUV images at $x=0$, which is shown as the vertical red line on the upper left panel. Another dataset for the south polar region is shown in Figure \ref{figure:map2} of Appendix A.

\begin{longtable}{ccccccccc}
\caption{Data sets} \label{table:datasets} \\
\toprule
No. & Date & start UT & end UT & Loc. & $N_x$ & $n_x$ & $\Delta t$ & $v_{\rm{offset}}$ \\
\midrule
\endfirsthead

\multicolumn{9}{c}%
{{\bfseries Table \thetable\ (continued)}} \\
\toprule
No. & Date & start UT & end UT & Loc. & $N_x$ & $n_x$ & $\Delta t$ & $v_{\rm{offset}}$ \\
\midrule
\endhead

\bottomrule
\multicolumn{9}{r}{{Continued on next page}} \\
\endfoot

\bottomrule
\endlastfoot
1 & 2007 Jan 14 & 16:05 & 18:58 & S & 2047 & 2047 & 4.8 & -12 \\
2 & 2007 Jan 15 & 16:08 & 19:02 & S & 2047 & 2047 & 4.8 & -22 \\
3 & 2007 Jan 18 & 12:15 & 13:42 & S & 1024 & 1024 & 4.8 & -34 \\
4 & 2007 Jan 18 & 15:40 & 17:09 & S & 1024 & 1024 & 4.8 & -33 \\
5 & 2007 Feb 16 & 02:20 & 03:45 & S & 1000 & 1000 & 4.8 & -7 \\
6 & 2007 Mar 7 & 12:34 & 13:06 & S & 380 & 380 & 4.8 & -19 \\
7 & 2007 Mar 16 & 12:02 & 14:55 & S & 2047 & 2047 & 4.8 & -15 \\
8 & 2007 Apr 7 & 18:14 & 21:07 & S & 2047 & 2047 & 4.8 & -19 \\
9 & 2007 Apr 8 & 18:33 & 19:31 & S & 689 & 689 & 4.8 & -5 \\
10 & 2007 Apr 15 & 10:53 & 11:52 & S & 690 & 690 & 4.8 & -21 \\
11 & 2007 Apr 21 & 18:27 & 20:18 & S & 512 & 512 & 12.8 & -16 \\
12 & 2007 Apr 22 & 06:15 & 09:09 & S & 2047 & 2047 & 4.8 & -35 \\
13 & 2007 Sep 1 & 20:35 & 01:11 & N & 2000 & 2000 & 8.0 & -18 \\
14 & 2007 Sep 6 & 00:28 & 05:57 & N & 2000 & 2000 & 9.6 & -16 \\
15 & 2007 Sep 8 & 01:07 & 06:36 & N & 2000 & 1801 & 9.6 & -18 \\
16 & 2007 Sep 9 & 13:05 & 15:49 & S & 1000 & 1000 & 9.6 & -45 \\
17 & 2007 Sep 10 & 01:15 & 06:39 & N & 1969 & 1969 & 9.6 & -18 \\
18 & 2007 Sep 11 & 14:06 & 19:36 & N & 2000 & 1951 & 9.6 & -31 \\
19 & 2007 Oct 13 & 11:19 & 11:51 & S & 380 & 380 & 4.8 & -35 \\
20 & 2008 Sep 20 & 10:12 & 17:38 & N & 1978 & 1351 & 12.8 & -43 \\
21 & 2008 Sep 24 & 10:37 & 17:43 & N & 1905 & 1601 & 12.8 & -36 \\
22 & 2009 Mar 16 & 10:17 & 17:04 & S & 1826 & 1051 & 12.8 & -22 \\
23 & 2009 Mar 20 & 08:01 & 13:51 & S & 1573 & 981 & 12.8 & -29 \\
24 & 2009 Sep 13 & 10:41 & 16:45 & N & 1579 & 1101 & 12.8 & -10 \\
25 & 2009 Sep 16 & 10:55 & 18:21 & N & 1972 & 1701 & 12.8 & -25 \\
26 & 2009 Sep 18 & 10:35 & 17:40 & N & 1854 & 1654 & 12.8 & -27 \\
27 & 2010 Mar 15 & 22:25 & 05:50 & S & 1753 & 811 & 12.8 & -39 \\
28 & 2010 Mar 20 & 11:13 & 17:29 & S & 1683 & 1231 & 12.8 & -34 \\
29 & 2010 Mar 21 & 10:07 & 17:44 & S & 1920 & 1171 & 12.8 & -28 \\
30 & 2010 Mar 24 & 10:50 & 16:49 & S & 1588 & 1121 & 12.8 & -28 \\
31 & 2010 Mar 26 & 09:42 & 17:04 & N & 1986 & 1001 & 12.8 & -14 \\
32 & 2010 Oct 1 & 10:05 & 17:30 & N & 1997 & 1151 & 12.8 & 2 \\
33 & 2010 Oct 2 & 10:46 & 18:09 & N & 1993 & 1551 & 12.8 & 1 \\
34 & 2010 Oct 4 & 10:24 & 17:39 & N & 1967 & 1561 & 12.8 & 16 \\
35 & 2010 Oct 6 & 10:06 & 15:59 & N & 1574 & 1001 & 12.8 & 12 \\
36 & 2011 Mar 24 & 11:04 & 17:09 & S & 1608 & 708 & 12.8 & -32 \\
37 & 2011 Oct 9 & 10:04 & 17:30 & N & 2014 & 1611 & 12.8 & -66 \\
38 & 2012 Mar 25 & 08:36 & 15:59 & S & 1864 & 901 & 12.8 & -33 \\
39 & 2012 Sep 8 & 11:12 & 18:38 & N & 2021 & 951 & 12.8 & 22 \\
40 & 2013 Mar 6 & 08:45 & 16:11 & S & 2041 & 1051 & 12.8 & -30 \\
41 & 2014 Mar 5 & 08:15 & 16:36 & S & 2009 & 1011 & 12.8 & -28 \\
42 & 2014 Sep 8 & 11:06 & 17:26 & N & 1625 & 651 & 12.8 & -3 \\
43 & 2015 Mar 4 & 10:36 & 17:59 & S & 2017 & 701 & 12.8 & -16 \\
44 & 2015 Sep 13 & 10:06 & 17:31 & N & 1979 & 901 & 12.8 & -38 \\
45 & 2016 Mar 6 & 10:04 & 17:29 & S & 2016 & 931 & 12.8 & -15 \\
46 & 2016 Sep 7 & 08:30 & 15:56 & N & 2018 & 1101 & 12.8 & -70 \\
47 & 2017 Mar 10 & 10:46 & 18:11 & S & 2017 & 1251 & 12.8 & -45 \\
48 & 2017 Sep 15 & 09:05 & 16:31 & N & 2004 & 1101 & 12.8 & -48 \\
49 & 2017 Sep 18 & 09:36 & 16:59 & S & 1995 & 995 & 12.8 & -25 \\
\end{longtable}
\vspace{-20pt}
\begin{tablenotes}
\item Notes.
\item {$N_x$~[pixel] : Number of slit positions before cutting out.}
\item {$n_x$~[pixel] : Number of slit positions used for the analysis.}
\item {$\Delta t$~[s] : Exposure time per slit position.}
\item {$v_{\rm{offset}}$~[m s$^{-1}$] : Velocity correction value.}
\end{tablenotes}

%%%%%%%%%%%%%%%%%% figure

%%%%%%%%% Section %%%%%%%%%  
\section{Method} \label{section:method} 
\begin{figure}[htbp]    %%%%%%%%%%%%%%%%%% offset
	\centerline{\includegraphics[width=1.0\textwidth,clip]{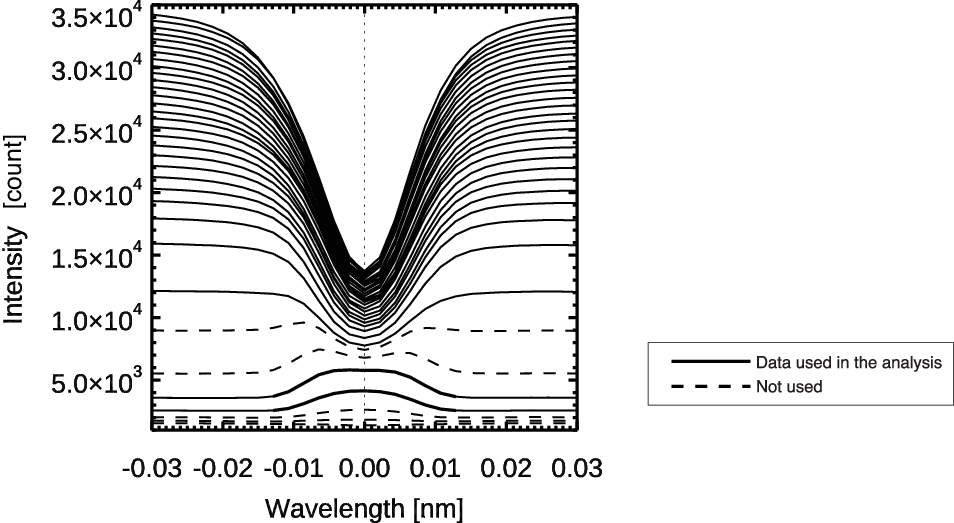}}
	\caption{Spatially-averaged spectral lines of Fe~{\sc i}~630.15~{nm} (dataset No. 13 in Table \ref{table:datasets}). The absorption lines are averaged from 63 to 89 degrees in one degree intervals. The emission lines are averaged from the limb to $1^{\prime\prime}.05$  above it, with an interval of $0^{\prime\prime}.15$ . The vertical and horizontal axes show the intensity and wavelength $- 630.15$ [nm], respectively. The solid lines show the data used in the analysis. The dashed lines indicate data not used in the analysis.}
    \label{figure:spectrum_abs_emi}
 \end{figure}

In this study, we use Doppler velocities derived from spatially averaged Stokes I profiles at each heliocentric angle. This approach is commonly adopted in previous studies (e.g. \citealp{lohner2019convective}). Appendix B compares Doppler velocities derived from averaged profiles with those obtained by averaging velocities derived at individual pixels. The difference between these two methods is relatively small compared with the overall dependence on heliocentric angles. We obtain spatially averaged absorption profiles observed on the solar disk at each heliocentric angle from $63^{\circ}$ to $89^{\circ}$ with a $1^{\circ}$ step size in the range of $\pm0.5^{\circ}$. The profiles observed above the limb are spatially averaged at an interval equivalent to one pixel ($0^{\prime\prime}.15$) from the limb position detected. Mixed profiles of absorption and emission lines are observed up to the second pixel from the limb position. The amplitude of the emission line at the fourth and fifth pixel is not sufficient for analysis, and scattered light is observed after the sixth pixel. We take the profiles at the second and third pixel from the limb as the emission line. Figure \ref{figure:spectrum_abs_emi} shows the variation of the spatially averaged profiles across the limb for dataset No.13. The solid lines indicate the absorption and emission profiles used in the analysis. The dashed lines show the data not used in the analysis.

%%%%%%%%%%%%%%%%%% figure
  
\begin{figure}[htbp]    %%%%%%%%%%%%%%%%%% spectrum
	\centerline{\includegraphics[width=1.0\textwidth,clip]{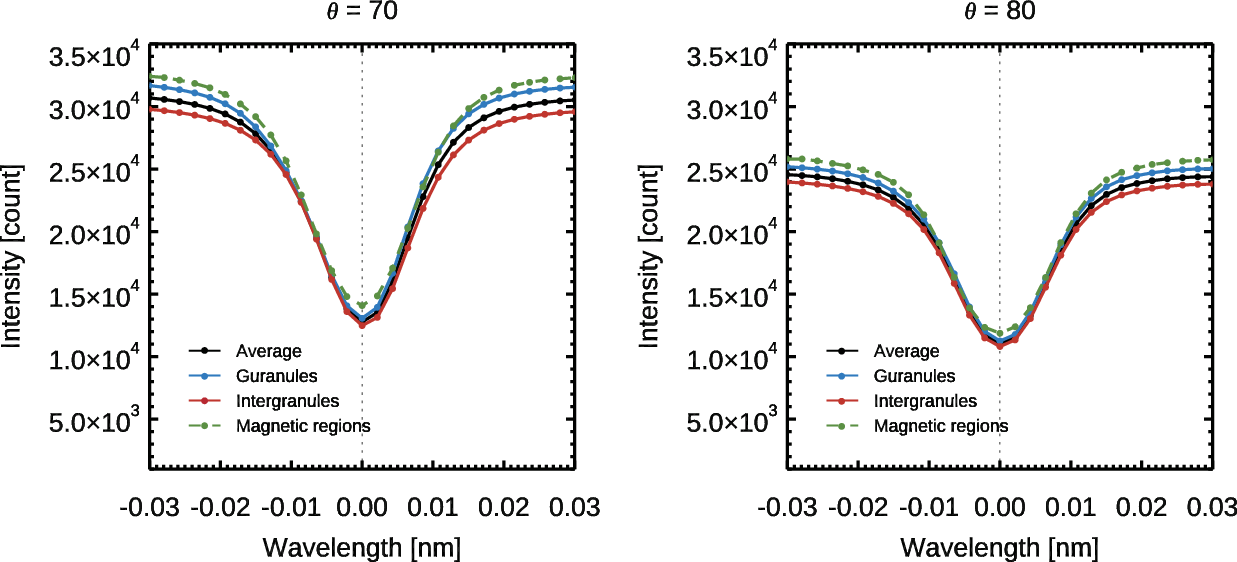}}
	\caption{Spatially-averaged spectral line at Fe~{\sc i}~$630.15$~{nm} after classification (dataset No. 13 in Table \ref{table:datasets}). We classify the absorption profiles into those from granules, intergranules, and magnetic regions using thresholds of the continuum intensities and the polarization degrees at each heliocentric angle. The left panel shows the spectral profiles at $\theta = 70^{\circ}$, and the right panel shows the spectral profiles at $\theta = 80^{\circ}$. The black lines represent the spectral profiles for the overall average. 
	The blue and red solid lines represent the spectral lines averaged over the areas of granule and intergranule regions, respectively, while the green dashed line represents the spectral lines averaged over the areas of the magnetic regions.}
    \label{figure:spectrum_70_80}
 \end{figure}

We classify the absorption profiles into those from granules, intergranules, and magnetic regions using thresholds of the total polarization degrees and the continuum intensity because the SOT/SP can resolve granule structures near the limb. The total polarization degree $\sqrt(Q^2 + U^2 + V^2)/I_{\rm{continuum}}$ is integrated over lines, which is processed using the SP\_PREP procedure \citep{lites2013spprep}. We set the threshold for total polarization at $0.6$\%. This $0.6$\% is three times larger than the total polarization degree averaged in internetwork areas. We define an absorption profile with a total polarization of $0.6$\% and more as `magnetic regions'. For absorption profiles with a total polarization of less than $0.6$\%, we further classify them using the continuum intensities. To account for the effect of limb darkening, we obtain the trend of the average continuum intensity as a function of the distance from the disk center by averaging the continuum intensity every $0^{\prime\prime}.3$. When the continuum intensity of an absorption profile at a spatial pixel is higher than the average, we define this absorption profile as a `granules'. The absorption profile whose continuum intensity is lower than the average is defined as an `intergranules'. Figure \ref{figure:spectrum_70_80} shows the spatially averaged profiles at $\theta = 70^{\circ}$ and $80^{\circ}$ in dataset No.13. The horizontal axis shows wavelengths from $630.15$~{nm}. The vertical axis shows intensity. The black solid lines show the spatially averaged profiles over the area, referred to as the ‘overall average’. The green dashed lines show the spatially averaged profiles of magnetic regions. The blue solid lines show the spatially averaged profiles of granules. The red solid lines show the spatially averaged profile of intergranules.

 %%%%%%%%%%%%%%%%%% figure

\begin{figure}[htbp]    %%%%%%%%%%%%%%%%%% filling factor
	\centerline{\includegraphics[width=1.0\textwidth,clip]{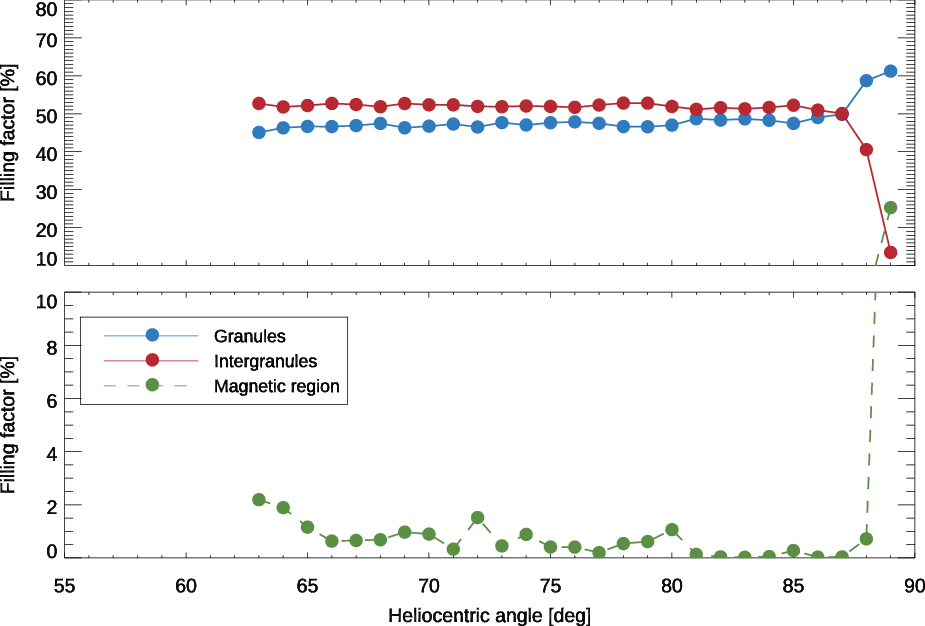}}
	\caption{Filling factor of granules (blue solid line), intergranules (red solid line), and  magnetic regions (green dashed line) as a function of the heliocentric angles $\theta$ for the data taken on 1 September 2007 (dataset No. 13 in Table\ref{table:datasets}).}
    \label{figure:fillingfactor}
 \end{figure}

The filling factor of each region varies with the heliocentric angle, as shown in Figure \ref{figure:fillingfactor}. The filling factor is determined as the ratio of the number of pixels in each region to the total number of pixels at each heliocentric angle. In Figure \ref{figure:fillingfactor}, the horizontal axis shows the heliocentric angle $\theta$ [$^{\circ}$], and the vertical axis shows the filling factor [\%]. The blue, red, and green lines indicate the granules, intergranules, and magnetic regions, respectively. The disk area is mostly covered by either granules or intergranules. The filling factor of the magnetic regions is small because the polar region is occupied by quiet regions. To characterize Doppler velocities of surface convection, we perform the analysis in this study for granules and intergranules and not for magnetic regions.

%%%%%%%%%%%%%%%%%% figure

To reveal the line-of-sight velocity gradient, we perform bisector analysis for absorption lines(e.g. \citealp{adam1976the}). The bisector is defined as the midpoints of the segment connecting the two ends at a specific intensity of the profile. In this study, the wavelengths corresponding to the specific intensity levels were obtained by linear interpolation between the observed data points. We determine the bisector position $\lambda_{\rm{bisector}}$ at a normalized intensity from $0.01$ to $0.99$ in steps of $0.01$, where the normalized intensity at linecore is $0$ and the continuum intensity is $1$. For the analysis of the Fe~{\sc i}~$630.15$~{nm} absorption line, we used $31$ spectral points centered on the line, covering the wavelength range from $-0.03225$~{nm} to $+0.03225$~{nm}. The wavelength at linecore $\lambda_{\rm{linecore}}$ is obtained using parabolic function of three points around the line minima. To avoid contamination from the line wings, we chose the minimal number of points necessary to isolate the linecore. For the emission lines, we determine the center wavelength $\lambda_{\rm{emission}}$ using the center of gravity method because the emission lines have a weak intensity and are often distorted in shape. The integration was performed over $-0.0129$~{nm} to $+0.0129$~{nm} around the nominal line center.  The Doppler velocities $v$ are obtained using
\begin{equation}
v = c\frac{\lambda_i - \lambda_0}{\lambda_0}, 
\end{equation}
where c is the speed of light, $\lambda_0$ is the wavelength of the spectral line (630.15~{nm}), and $\lambda_{i}$ is the observed wavelength such as $\lambda_{\rm{biector}}$, $\lambda_{\rm{linecore}}$, or $\lambda_{\rm{emission}}$. \cite{ishikawa2020study} estimated the uncertainty in bisector analysis to be approximately $40$~{m s$^{-1}$} near the linecore, based on the measurement error of intensities at each pixel of the SOT/SP data. Since the current analysis uses the averaged profiles from many pixels, the statistical uncertainty of the bisector velocity is sufficiently small.

Although wavelength calibration of the SOT/SP data is performed in the standard calibration routine SP\_PREP \citep{lites2013spprep}, the SP\_PREP routine may not have been able to correctly determine the zero point of the wavelength. This is because the zero point is set so that the line center of the spatially averaged spectral line becomes $630.15$~{nm}, without considering the redshift near the solar limb. In this study, we perform a velocity correction using the accurate spectroscopic measurements of the Fe~{\sc i} $630.15$~{nm} line obtained with a laser frequency comb \citep{lohner2019convective} to adjust the zero point of each dataset. We performed a linear approximation of the dependence of overall average velocity on the heliocentric angle below $\theta = 80^{\circ}$ obtained from the analysis of the Hinode observations in this study: 
\begin{equation}
v = a \theta + b,  
\end{equation}
where $a$ and $b$ are obtained from linear fits to each dataset. The correction value $v_{\rm{offset}}$ is determined so that the velocity at $\theta = 66.4^{\circ}$ is $18$~{m~s$^{-1}$} given by \citet{lohner2019convective}, 
\begin{equation}
\label{equation:offset}
v_{\rm{offset}}= 66.4 a  + b - 18.
\end{equation}

\begin{figure}[htbp]    %%%%%%%%%%%%%%%%%% spectrum
	\centerline{\includegraphics[width=1.0\textwidth,clip]{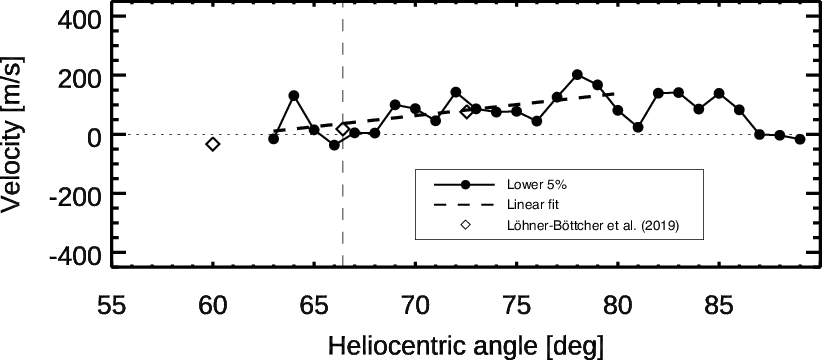}}
	\caption{Doppler velocity before velocity correction for the data taken on 1 September 2007 (dataset No.13 in Table \ref{table:datasets}). The filled circle symbols and solid line show the bisector velocities of the lower 5\% of the averaged profiles as a function of the heliocentric angles $\theta$ [$^{\circ}$]. The dashed line shows the result of the linear fitting of the bisector velocities between $\theta = 63^{\circ}$ and $80^{\circ}$. The diamond symbols show the bisector velocity of the lower $5$\% at $\mu = \cos\theta=0.3, 0.4,$ and $0.5$ obtained by \citet{lohner2019convective}. See Figure \ref{figure:offset2} for another dataset taken in the south polar region.}
    \label{figure:offset}
 \end{figure}

In Figure \ref{figure:offset}, the diamonds show the Doppler velocities at Fe~{\sc i}~630.15~{nm} obtained by \citet{lohner2019convective}, $-33$~{m s$^{-1}$} at $\theta = 60^{\circ}$, $18$~{m s$^{-1}$} at $\theta = 66.4^{\circ}$, and 76~{m s$^{-1}$} at $\theta = 72.5^{\circ}$. The solid line shows the Doppler velocities $V_{\rm{lower5\%}}$, which is the average of the bisector velocities of lower $5$\% and line core velocity using dataset No.13. The dashed line shows a linear fitting of $V_{\rm{lower5\%}}$ up to $\theta = 80^{\circ}$. We determine the velocity offset for each dataset, as shown in Table \ref{table:datasets}. Although the SP\_PREP calibration routine corrects for instrumental effects such as wavelength drift and flat field, small residuals may persist and cause uncertainties in the velocity measurements. Moreover, Doppler velocities from supergranular motions may not fully average out within each dataset, leading to differences in the velocity profiles across datasets. The average of standard deviation among the datasets is $67$~{m s$^{-1}$}, reflecting the combined influence of these error sources.

  %%%%%%%%%%%%%%%%%% figure

%%%%%%%%% Section %%%%%%%%%  
\section{Results}\label{section:results} 
\subsection{Variation in the Doppler velocities}
\begin{figure}[htbp]  %%%%%%%%%%%%%%%%%% v
   \centerline{\includegraphics[width=1.0\textwidth,clip]{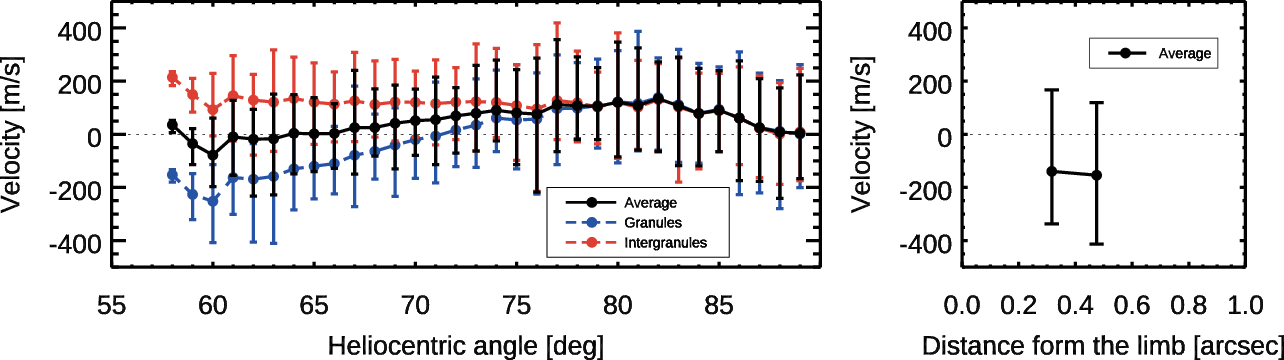}}
   \caption{Averaged Doppler velocity for 49 datasets. The panel on the left shows the Doppler velocities for the lower 5\% of the absorption line as a function of angle. The vertical axis is the Doppler velocity [m~s$^{-1}$], and the horizontal axis is the heliocentric angle $\theta$ [$^{\circ}$]. 
   The black solid line shows the velocity of the overall average, while the blue and red dashed lines show the velocities of the granules and intergranules, respectively. The error bars show the maximum and minimum velocities of the datasets. The right panel shows the Doppler velocity of the emission lines as a function of distance from the limb. The vertical axis is the Doppler velocity [m~s$^{-1}$], and the horizontal axis is the distance from the limb [arcsec]. }
    \label{figure:v}
 \end{figure}

Figure \ref{figure:v} shows Doppler velocities of Fe~{\sc i}~$630.15$~{nm}. The left panels show Doppler velocities of the absorption lines observed on the disk. The horizontal axis shows heliocentric angles $\theta$ [$^{\circ}$]. The vertical axis shows Doppler velocities [m~s$^{-1}$]. Positive and negative velocities correspond to redshifts and blueshifts, respectively. The black solid line shows the Doppler velocities of the overall average, which is the average over the region of each heliocentric angle. The blue dashed line shows the Doppler velocities of granules averaged over the bright and non-magnetic regions. The red dashed line shows the Doppler velocities of intergranules averaged over the dark and non-magnetic regions. The error bars indicate the maximum and minimum velocities among the datasets. The right panel of Figure \ref{figure:v} shows Doppler velocities of the emission lines of Fe~{\sc i}~$630.15$~{nm} observed within $1^{\prime\prime}$ from the limb. The vertical axis of the right panel shows Doppler velocities [m~s$^{-1}$].  The horizontal axis shows the distance from the limb [arcsec]. The circle symbols indicate the mean of all datasets. The error bars indicate the maximum and minimum velocities among the datasets.

%%%%%%%%%%%%%%%%%% figure

As shown in the left panel, the Doppler velocities of the overall average shown by the black solid line increase in the redshift from $\theta = 65^{\circ}$ to $80^{\circ}$. The variation in Doppler velocities becomes small above $\theta = 80^{\circ}$. The Doppler velocities get close to zero above $\theta = 85^{\circ}$. The Doppler velocities of the granules shown by the blue dashed line increase in $240$~{m~s$^{-1}$} from $\theta = 65^{\circ}$ to $80^{\circ}$. The Doppler velocities become $120$~{m~s$^{-1}$} at $\theta = 80^{\circ}$. The Doppler velocities of the intergranules remain nearly constant at around $120$~{m~s$^{-1}$} in the heliocentric angle range between $\theta = 65^{\circ}$ and $80^{\circ}$, varying within $\pm$20~{m~s$^{-1}$}. The velocity variation of the intergranules is smaller than that of the granules. As shown in the right panel, the emission lines above the limb are more blue-shifted than the absorption lines on the disk near the limb. The average Doppler velocity of the emission lines is $-130$~{m~s$^{-1}$}.

\subsection{Variation in the line bisectors}
\begin{figure}[htbp] %%%%%%%%%%%%%%%%%% bisector
   \centerline{\includegraphics[width=1.0\textwidth,clip]{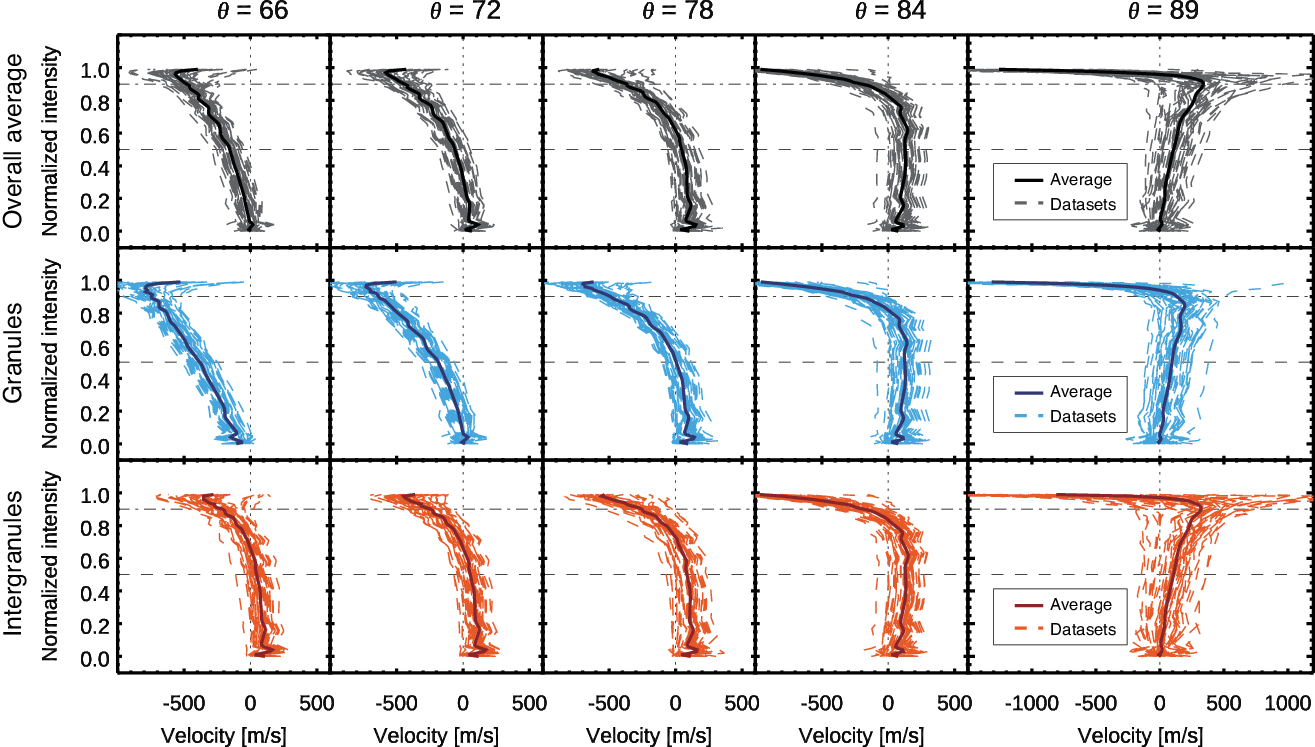}}
   \caption{The bisector of spatially-averaged absorption profiles of the Fe~{\sc i}~630.15~{nm} line. The vertical axis shows the normalized intensity, and the horizontal axis shows the Doppler velocity [m~s$^{-1}$]. 
 From left to right, the panels show the bisectors at at $\theta = 66, 72, 78, 84,$ and $89^{\circ}$. The solid lines indicate the bisector averaged for all data, and the dashed lines show the bisector for each dataset. The top, middle, and bottom panels show the bisectors of the overall average, granules, and intergranules, respectively.}
    \label{figure:bisector}
 \end{figure}

 Figure \ref{figure:bisector} shows bisector velocities at $\theta = 66, 72, 78, 84,$ and $89^{\circ}$. The panels show the bisector of the overall average (black), granules (blue), and intergranules (red), from the top row to bottom row. The columns show the bisector at $\theta = 66, 72, 78, 84,$ and $89^{\circ}$ from left to right. The horizontal axis shows the Doppler velocity [m~s$^{-1}$], and the vertical axis shows normalized intensity. The results of individual datasets are shown by the dashed lines. The means of all datasets are shown as solid lines. As shown in the top row, the bisector of the overall average at $\theta = 66^{\circ}$ shows the \textbackslash-shape. It becomes an I-shape toward the limb. In the middle row, the bisector of granules at $\theta = 66^{\circ}$ has clearly a velocity gradient compared with that of the overall average. In the bottom panel, the bisector of intergranules keeps the I-shape from $66$ to $89^{\circ}$.
 
 %%%%%%%%%%%%%%%%%% figure

\begin{figure}[htbp]  %%%%%%%%%%%%%%%%%% v + bisector
   \centerline{\includegraphics[width=1.0\textwidth,clip]{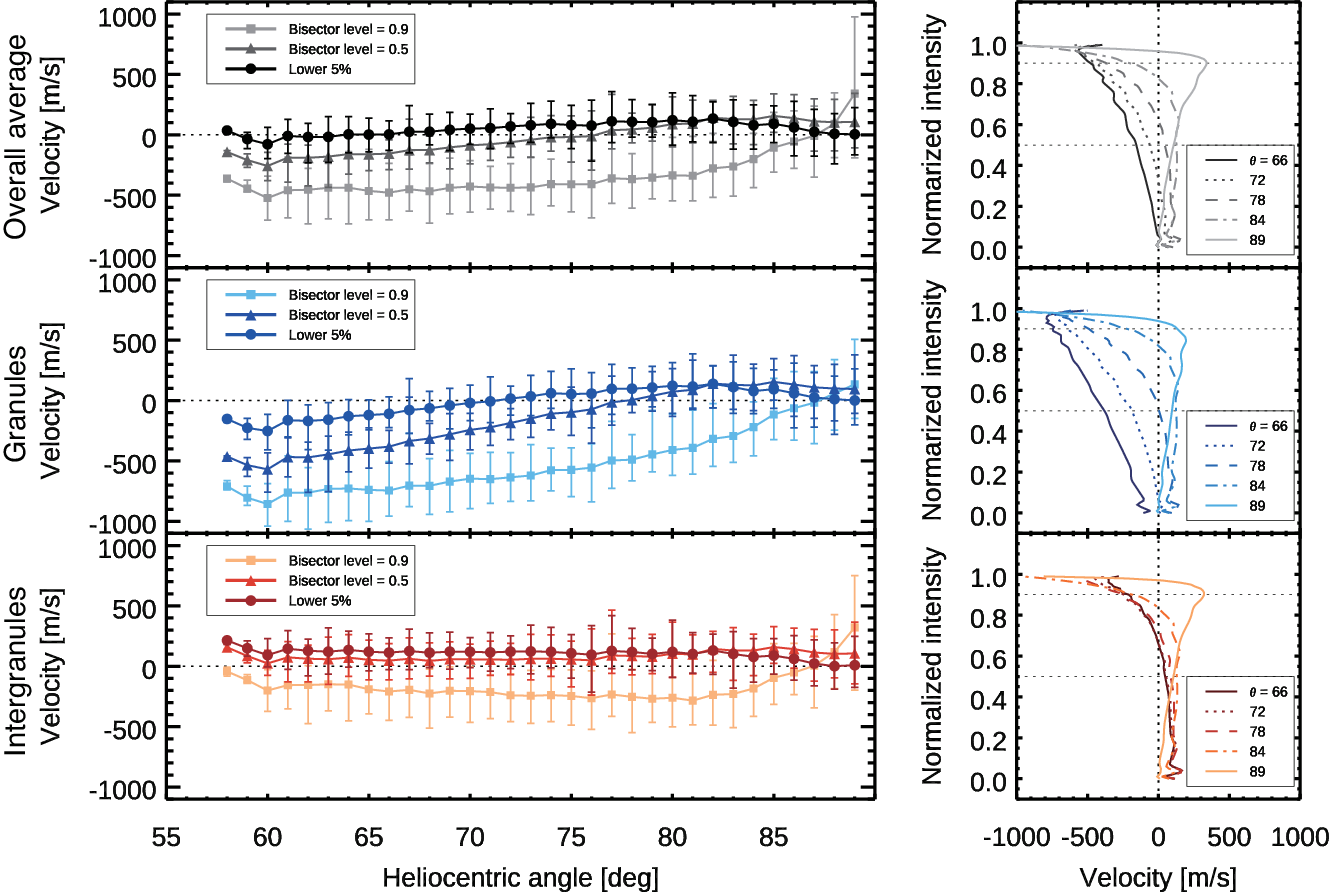}}
   \caption{Left: Doppler velocities at three bisector levels as a function of heliocentric angles. The vertical axis shows the Doppler velocity [m~s$^{-1}$] and the horizontal axis shows the angle [$^{\circ}$]. The upper panel shows the results using the profiles of the overall average (black), the middle panel for granules (blue), and the lower panel for intergranules (red). The Doppler velocities at normalized intensity of $0.9$ are shown by squares, the Doppler velocities at normalized intensity of $0.5$ by the triangles, and the Doppler velocities of lower $5$\% by the circles. Right: bisectors obtained at several heliocentric angles. The vertical axis shows normalized intensity, and the horizontal axis shows the Doppler velocity [m~s$^{-1}$]. From darkest to lightest, the colored lines show the bisector at $\theta =66, 72, 78, 84,$ and $89^{\circ}$, corresponding to the average bisectors shown by the thick solid lines in Figure \ref{figure:bisector}.
}
    \label{figure:v+bisector}
 \end{figure}

The left column of Figure~\ref{figure:v+bisector} shows the Doppler velocities derived by bisectors at 0.90 and 0.05 and for the lower $5$\%. The vertical axis shows the Doppler velocity [m~s$^{-1}$] and horizontal axis shows the heliocentric angle [$^{\circ}$]. The upper, middle, and lower panels show the results using the spatially-averaged spectral lines for the overall average (black), granules (blue), and intergranules (red), respectively. The Doppler velocities at a normalized intensity of 0.90 are indicated by the squares. The Doppler velocities at a normalized intensity of 0.50 are indicated by the triangles, and the Doppler velocities of the lower 5\% are indicated by the circles. The error bars show the maximum and minimum velocities of the datasets. The right column of Figure~\ref{figure:v+bisector} shows the bisector at $\theta = 66, 72, 78, 84,$ and $89^{\circ}$, which are shown by the thick solid lines in Figure \ref{figure:bisector}. The bisector velocities of the overall average are more red-shifted around the line core than near the wing. The velocity differences between the line core and wing decrease toward the limb, and bisector velocities of the wing are red-shifted after $\theta = 88^{\circ}$. The bisector velocities of the granules have a larger velocity difference and show more redshift around the line core than near the wing. The velocity difference decreases after $\theta = 80^{\circ}$. The velocity difference between the line core and wing of intergranules is smaller than that of granules.
  %%%%%%%%%%%%%%%%%% figure

 %%%%%%%%% Section %%%%%%%%%  
 \section{Discussion}\label{section:discussion} 
 \subsection{Comparison with previous observations}
The Doppler velocities of the overall average exhibited an increase in redshift toward $\theta = 75^{\circ}$. The increase from $\theta = 66^{\circ}$ to $72^{\circ}$ was $70$ m~s$^{-1}$, which was consistent with the observation result of \citet{lohner2019convective}. The overall average Doppler velocities were almost constant around $\theta = 80^{\circ}$. The Doppler velocities of the overall average exhibited a decrease in redshift above $\theta = 85^{\circ}$ and became nearly zero at $\theta = 89^{\circ}$. This can be explained by averaging over the randomly distributed horizontal flows, which reduces the correlation between the Doppler velocity and continuum intensity.

The Doppler velocities of granules showed a decrease in blueshift toward $\theta = 80^{\circ}$, as shown in Figures \ref{figure:v} and \ref{figure:v+bisector}. The increase from $\theta = 64^{\circ}$ to $74^{\circ}$ was $190$~{m~s$^{-1}$}. This result suggests that the variation in the overall average Doppler velocities is primarily due to the bright granules. In this study, we defined granules as regions where the continuum intensity was larger than the average continuum intensity. The observed decrease in the blueshift of granules is consistent with the interpretation that redshifted horizontal flows become brighter and thus more significant toward the limb, which has been qualitatively considered as one of the reasons for the increase in redshift toward the limb \citep{beckers1978some, cegla2018stellar}.
The Doppler velocities of the intergranules remained nearly constant at around $120$~{m~s$^{-1}$} in the heliocentric angle range between $\theta = 65^{\circ}$ and $80^{\circ}$, varying within $\pm$20~{m~s$^{-1}$}. 

The bisector exhibited a \textbackslash -shape at $\theta = 66^{\circ}$, consistent with the observation reported by \citet{lohner2019convective}. At $\theta = 78^{\circ}$, the upper part of the bisector showed a \textbackslash -shape, whereas the lower part became an I-shape. Its bisector was consistent with the observations of \citet{adam1976the} and \citet{dravins1982photospheric}. 
The bisector became an I-shape at $\theta = 84^{\circ}$. This I-shape indicates that the velocity difference in the line-of-sight direction is small. These observations can be interpreted as suggesting that the geometrical thickness of the line-forming region decreases toward the limb. The bisectors of granules at $\theta = 66^{\circ}$ showed a clear velocity difference compared with intergranules. The bisectors of intergranules were almost symmetric. These suggest that the geometrical thickness of the atmosphere in granules may be larger than that in intergranules.

The Doppler velocities above $\theta = 76^{\circ}$ in Figure \ref{figure:v} and bisectors after $\theta = 78^{\circ}$ in Figure \ref{figure:bisector} showed similar trends for the granules and intergranules. This is because a spatial resolution of $0^{\prime\prime}.3$ corresponds to about $1.3$~{Mm} on the solar surface at $\theta = 76^{\circ}$, making it impossible to isolate granules and intergranules.

At $\theta = 89^{\circ}$, the average velocity of the absorption line was $-20$~{m~s$^{-1}$} with a standard deviation of $70$~{m~s$^{-1}$}. The average velocities of the emission lines were $-130$ and $-140$~{m~s$^{-1}$} with standard deviations of $100$ and $110$~{m~s$^{-1}$}, respectively. The velocity of the absorption line was approximately one standard deviation away from the velocity of the emission lines. Although the difference between the velocities of emission lines and absorption lines was not statistically significant, the emission lines tended to exhibit a blue shift. The absolute velocity of the emission lines could be approximately zero if the emission profiles are created by line-of-sight integration in the upper photosphere under optically thin conditions. These blueshifted emission lines can be partially due to insufficient velocity correction. The velocity correction in this study was done by the linear approximation of the velocity trend below $80^{\circ}$, and the correction value was set so that the velocity at $66^{\circ}$ was $18$~{m~s$^{-1}$}, as given by \citet{lohner2019convective}. The slope of the linear approximation in this study did not perfectly match the amount of change in \citet{lohner2019convective}. This discrepancy may arise because the linear approximation was too simple to capture the range where the velocity trend changes (around $\theta = 80^{\circ}$). Another possibility is the difference in the observing fields-of-view between this study and the previous one. We cannot determine whether the velocity correction is sufficiently accurate or whether the emission line velocity truly represents zero. If we use emission lines as the zero-velocity reference, the absorption line may be redshifted by about $-130$~{m~s$^{-1}$}.

 \subsection{Comparison with numerical simulations}
This section compares our results with numerical simulations of a photospheric spectral line. Spectral line profiles near the solar limb can be synthesized by solving radiative transfer in a tilted box from a numerical simulation. However, near the limb, the optical path becomes much longer, making accurate synthesis difficult. Furthermore, because the line-of-sight passes through the upper photosphere, realistic modeling requires 3D simulations that extend to the upper photosphere. Several studies have attempted to synthesize spectral lines near the solar limb. For example, \citet{cegla2018stellar} employed a 3D magnetohydrodynamic solar simulation MURaM with the Non-local thermodynamic equilibrium Inversion COde using the Lorien Engine (NICOLE) to calculate Doppler velocities and bisectors up to $\theta = 80^{\circ}$, focusing on granules and intergranules. \citet{delacruzrodriguez2011solar} performed synthesis using 3D hydrodynamic simulation with the multi-level non-LTE radiative transfer code (MULTI) up to $\mu=\cos\theta=0.3$.

In this study, both the overall average and granular Doppler velocities exhibited decreases in blueshift from $\theta = 66^{\circ}$ to $72^{\circ}$, by approximately $70$ m~s$^{-1}$ and $190$~{m~s$^{-1}$} respectively. These trends are consistent with the results of numerical simulations \citep{cegla2018stellar}. In contrast, the Doppler velocities in intergranules showed no significant variation in our result, which is inconsistent with results of \citet{cegla2018stellar} that show a decrease in blueshift. A possible reason for this discrepancy is the limited spatial resolution of the observations. Although the nominal resolution of Hinode SOT/SP is approximately $0^{\prime\prime}.3$, corresponding to about 600~km at $\theta = 60^{\circ}$, the effective resolution degrades at larger heliocentric angles. Since intergranules are typically 300~km wide, spatial blending of granules and intergranules likely smoothed out their velocity features. This may explain the flattened trend of the intergranule velocities toward the limb.

At $\theta = 66^{\circ}$, the bisector of the overall average in this study exhibited a \textbackslash -shape, which is consistent with the results of numerical simulations \citep{cegla2018stellar, delacruzrodriguez2011solar}. The velocity difference between linecore and wing in the bisector is approximately $500$~{m~s$^{-1}$}, which is consistent with \citet{delacruzrodriguez2011solar}, whereas \citet{cegla2018stellar} showed a smaller value of about $200$~{m~s$^{-1}$}. The main cause of this difference likely arises from the inclusion of magnetic fields. \citet{delacruzrodriguez2011solar} used non-magnetic simulations, while \citet{cegla2018stellar} incorporated magnetic fields. In this study, we analyzed a quiet-Sun region, where the area with magnetic regions occupies less than 1\% at $\theta = 66^{\circ}$, and thus our results are in closer agreement with the results of the non-magnetic simulations.

In contrast, there are few previous studies on the synthesis of the off-limb profiles of a photospheric line. One example is \citet{shelyag2015spectro}, who synthesized Stokes profiles of the Fe~{\sc i}~630.15~{nm}  and Fe~{\sc i}~630.25~{nm} lines above the solar limb and reproduced their transition from absorption to emission. While he obtained the averaged emission profiles, the net Doppler shift was not studied in detail. Our observations indicate that, on average, the emission lines show blueshifts relative to the nearby absorption lines within the disk. Further studies are needed to clarify whether this trend reflects intrinsic atmospheric dynamics or observational effects such as a calibration offset. Future numerical simulations that reproduce the observing geometry would further improve our understanding of the physical processes responsible for Doppler-shift behaviour in spectral lines near and beyond the limb.

To improve quantitative comparison with numerical simulations,  the following improvements are needed: 1) Improvement of spatial and temporal resolution. Using an instrument with better spatial resolution than the Hinode SOT/SP, such as Daniel K. Inouye Solar Telescope (DKIST) Visible Spectro-Polarimeter (ViSP; \citealp{dewijn2022instrumentation}),  would allow granules and intergranules to be resolved up to the limb. High-resolution observations would allow us to track how velocity fields and bisector shapes evolve dynamically near the limb and to compare them directly with high-resolution simulations. 2) More accurate absolute velocity measurements. One approach is to combine high spatial resolution with precise velocity calibration using narrow telluric lines or a laser frequency comb within a single instrument. 3) Multi-wavelength observations. This study used only the Fe~{\sc i} 630.15~{nm} line, which provides information for the photosphere. Other spectral lines can resolve velocity fields at different altitudes from the deep photosphere to the low chromosphere, as demonstrated by observations with the Sunrise Chromospheric Infrared spectro-Polarimeter (SCIP; \citealp{katsukawa2020sunrise}).

%% Figure 
%
% \begin{figure} 
% \centerline{\includegraphics[width=0.5\textwidth,clip=]{<fig.eps>}}
% \caption{}%\label{fig:?}
% \end{figure}

%% Table
%
% \begin{table}
% \caption{}%\label{tbl:?}
% \begin{tabular}{}     
% \hline
% \multicolumn{2}{c}{<>}
% <data>
% \hline
% \end{tabular}
% \end{table}
  
%%%%%%%%%%%%%%%%%%%%%%%%%%%%%%%%%%%%%%%%%%%%%%%%%%%%%%%%%%%%%%%%%%%%%%%%%%%
%% Appendix
%
\appendix

\section{Results from the south limb}
\begin{figure}[htbp]    %%%%%%%%%%%%%%%%%% MAP
   \centerline{\includegraphics[width=1.0\textwidth,clip]{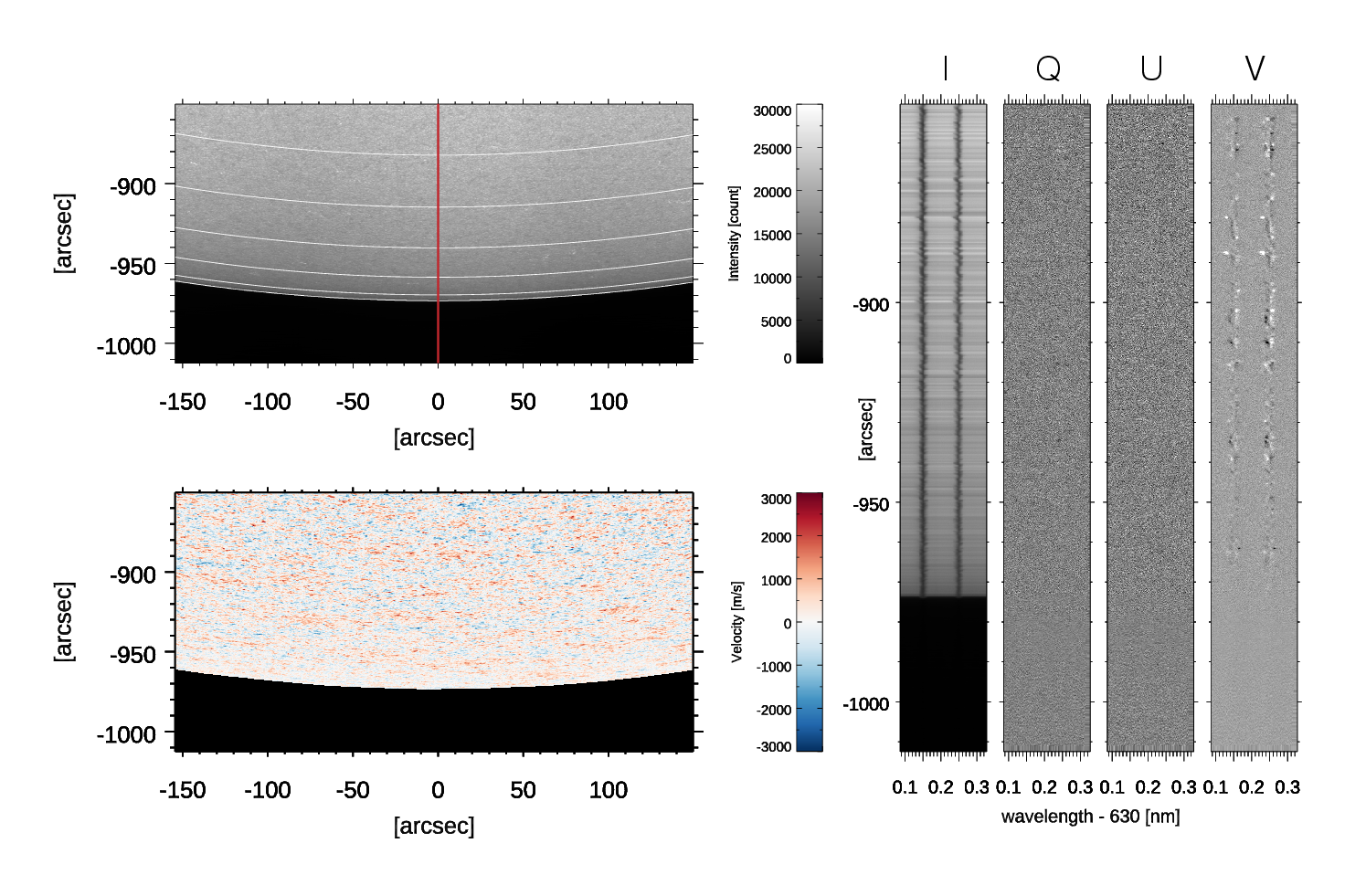}}
 \caption{Same as Figure \ref{figure:map} but for the south polar region (dataset No. 1 in Table \ref{table:datasets}).
}
    \label{figure:map2}
 \end{figure}

\begin{figure}[htbp]    %%%%%%%%%%%%%%%%%% spectrum
	\centerline{\includegraphics[width=1.0\textwidth,clip]{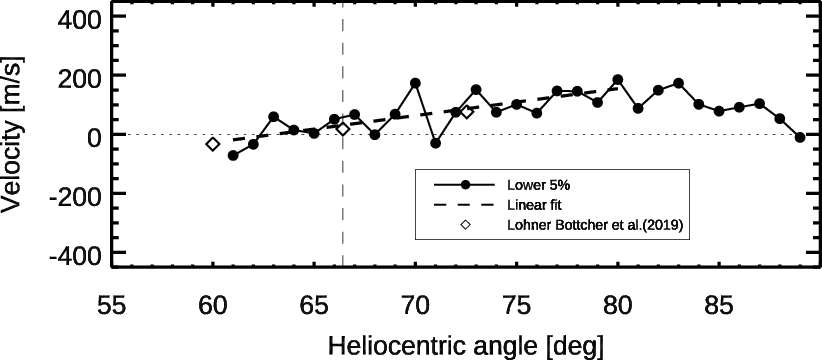}}
	\caption{Same as Figure \ref{figure:offset} but for the south polar region (dataset No. 1 in Table \ref{table:datasets}).} 
\label{figure:offset2}
 \end{figure}

We here present an example from the south polar region taken on 2007 January 14 (dataset No. 1 in Table \ref{table:datasets}) in Figures \ref{figure:map2} and \ref{figure:offset2}.

%%%%%%%%%%%%%%%%%% figure
  %%%%%%%%%%%%%%%%%% figure

\section{Comparison of Doppler velocities from two averaging methods}
We compare Doppler velocities derived from spatially averaged profiles with those obtained by averaging velocities calculated individually at each pixel as shown in Figure 11. The vertical axis shows the Doppler velocity [m s$^{-1}$], and the horizontal axis shows heliocentric angle $\theta$ [$^{\circ}$]. The solid lines indicate the Doppler velocities derived from the spatially averaged profiles, and the dotted lines indicate those obtained by averaging the velocities calculated individually at each pixel. The Doppler velocities at a normalized intensity of 0.90 are indicated by the asterisk. The Doppler velocities at a normalized intensity of 0.50 are indicated by the open circle, and the Doppler velocities of the linecore are indicated by the filled circle. The difference between the two methods is small. Although slightly larger differences are seen in the wing and near the limb ($\theta = 89^{\circ}$), most differences remain within 20 to 60~{m s$^{-1}$}.

 %%%%%%%%%%%%%%%%%% figure
\begin{figure}[htbp]  %%%%%%%%%%%%%%%%%% v + bisector
   \centerline{\includegraphics[width=1.0\textwidth,clip]{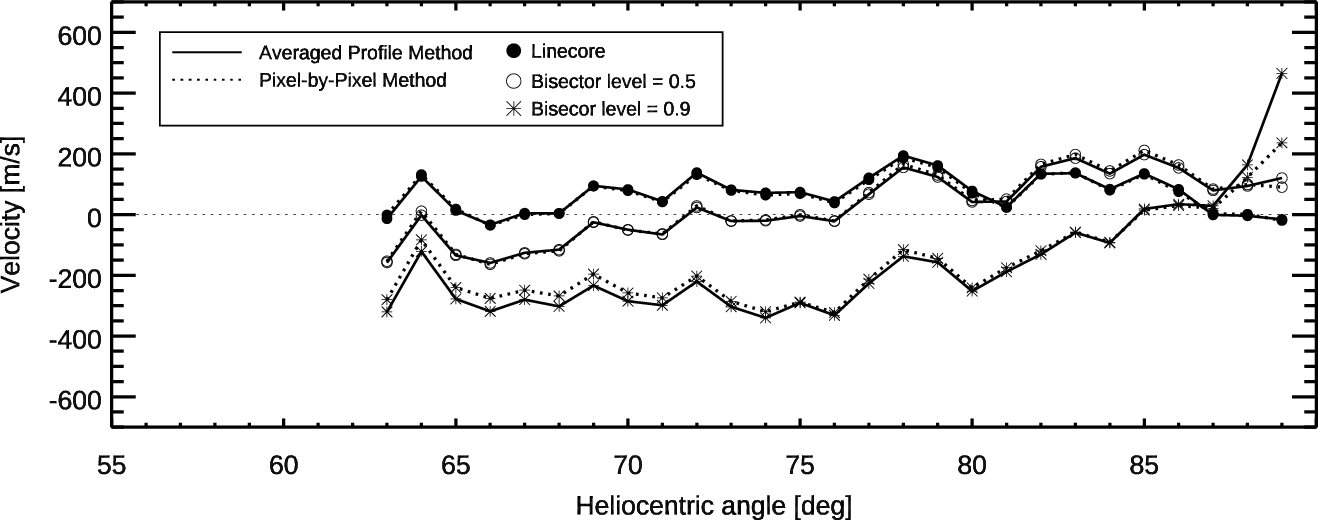}}
   \caption{Doppler velocities for the dataset taken on 1 September 2007 (Dataset No.13 in Table \ref{table:datasets}). The solid line represents the velocities derived from the averaged profiles, while the  dotted line indicates those obtained using the pixel-by-pixel approach.}
    \label{figure:v_comparison}
 \end{figure}

%%%%%%%%%%%%%%%%%%%%%%%%%%%%%%%%%%%%%%%%%%%%%%%%%%%%%%%%%%%%%%%%%%%%%%%%%%%
%% Acknowledgements
%
\begin{acks}
Hinode is a Japanese mission developed and launched by ISAS/JAXA, collaborating with NAOJ as a domestic partner, NASA and STFC (UK) as international partners. Scientific operation of the Hinode mission is conducted by the Hinode science team organized at ISAS/JAXA. This team mainly consists of scientists from institutes in the partner countries. Support for the post-launch operation is provided by JAXA and NAOJ (Japan), STFC (U.K.), NASA, ESA, and NSC (Norway). This work was supported by the JSPS KAKENHI Grant Number JP18H05234, JP20KK0072, JP23H01220, and JP23K25916. This work was supported in part by Japan Foundation for Promotion of Astronomy.
\end{acks}

\begin{ethics}
\begin{conflict}
The authors declare that they have no conflicts of interest.
\end{conflict}
\end{ethics}

%% Available additional data environments:
%% required: authorcontribution, fundinginformation, dataavailability
%% optional: materialsavailability, codeavailability
% \begin{authorcontribution}
%
% \end{authorcontribution}
%
% \begin{fundinginformation}
%
% \end{fundinginformation}
%
% \begin{dataavailability}
%
% \end{dataavailability}
%
% \begin{ethics}
% \begin{conflict}
%
% \end{conflict}
% \end{ethics}

%%% %%%%%%%%%%%%%%%%%%%%%%%%%%%%%%%%%%%%%%%%%%%%%%%%%%%%%%%%%%%
%% Bibliography
%
% Using BibTeX
%
\bibliographystyle{spr-mp-sola}
\bibliography{sola_bibliography_example}  
%
% Without BibTeX 
% \begin{thebibliography}{}
% \bibitem[\protect\citeauthoryear{Author}{Year}]{key}
%   <bibliographical entry>
%
% \bibitem[\protect\citeauthoryear{}{}]{}
%   
%  
% \end{thebibliography}

\end{document}